\documentclass[12pt,preprint]{aastex}
\usepackage{epsf}
\newcommand{\Msun}{\mathrm{M}_{\odot}}

\begin{document}

\shortauthors{Prieto et al.}

\title{LBT Discovery of a Yellow Supergiant Eclipsing Binary in the
Dwarf Galaxy Holmberg~IX\altaffilmark{1}}

\author{J.~L.~Prieto\altaffilmark{2}, K.~Z.~Stanek\altaffilmark{2},
C.~S.~Kochanek\altaffilmark{2}, D.~R.~Weisz\altaffilmark{3},
A.~Baruffolo\altaffilmark{4}, J.~Bechtold\altaffilmark{5},
V.~Burwitz\altaffilmark{6}, C.~De~Santis\altaffilmark{7},
S.~Gallozzi\altaffilmark{7}, P.~M.~Garnavich\altaffilmark{8},
E.~Giallongo\altaffilmark{7}, J.~M.~Hill\altaffilmark{9},
R.~W.~Pogge\altaffilmark{2}, R.~Ragazzoni\altaffilmark{4},
R.~Speziali\altaffilmark{7}, D.~J.~Thompson\altaffilmark{9},
R.~M.~Wagner\altaffilmark{2,9}}

\altaffiltext{1}{Based on data acquired using the Large Binocular
Telescope (LBT). The LBT is an international collaboration among
institutions in the United States, Italy and Germany. LBT Corporation
partners are: The University of Arizona on behalf of the Arizona
university system; Istituto Nazionale di Astrofisica, Italy; LBT
Beteiligungsgesellschaft, Germany, representing the Max-Planck Society,
the Astrophysical Institute Potsdam, and Heidelberg University; The Ohio
State University, and The Research Corporation, on behalf of The
University of Notre Dame, University of Minnesota and University of
Virginia.}

\altaffiltext{2}{\small Department of Astronomy, Ohio State University, Columbus, OH 43210}
\altaffiltext{3}{\small Department of Astronomy, University of Minnesota, Minneapolis, MN 55455}
\altaffiltext{4}{\small INAF, Osservatorio Astronomico di Padova, vicolo dell'Osservatorio 5, I-35122 Padova, Italy}
\altaffiltext{5}{\small Steward Observatory, The University of Arizona, Tucson, AZ 85721}
\altaffiltext{6}{\small Max-Planck-Institut f\"ur extraterrestrische Physik, Giessenbachstra\ss e, 85741 Garching, Germany}
\altaffiltext{7}{\small INAF, Osservatorio Astronomico di Roma, via di Frascati 33, I-00040 Monteporzio, Italy}
\altaffiltext{8}{\small University of Notre Dame, 225 Nieuwland Science, Notre Dame, IN 46556-5670}
\altaffiltext{9}{\small Large Binocular Telescope Observatory, University of Arizona, 933 N. Cherry Ave., Tucson, AZ  85721-0065}

\email{prieto@astronomy.ohio-state.edu}

\begin{abstract}
In a variability survey of M81 using the Large Binocular Telescope we
have discovered a peculiar eclipsing binary ($M_{V} \sim -7.1$) in the
field of the dwarf galaxy Holmberg~IX. It has a period of 271~days and
the light curve is well-fit by an overcontact model in which both stars
are overflowing their Roche lobes. It is composed of two yellow
supergiants ($V-I \simeq 1$~mag, $T_{\rm eff}\simeq 4800$~K), rather
than the far more common red or blue supergiants. Such systems must be
rare. While we failed to find any similar systems in the literature, we
did, however note a second example. The SMC F0 supergiant R47 is a
bright ($M_{V}\sim -7.5$) periodic variable whose All Sky Automated
Survey (ASAS) light curve is well-fit as a contact binary with a 181~day
period. We propose that these systems are the progenitors of supernovae
like SN~2004et and SN~2006ov, which appeared to have yellow
progenitors. The binary interactions (mass transfer, mass loss) limit
the size of the supergiant to give it a higher surface temperature than
an isolated star at the same core evolutionary stage. We also discuss
the possibility of this variable being a long-period Cepheid.
\end{abstract}



\section{Introduction}

Although small in number, massive stars are critical to the formation
and evolution of galaxies. They shape the ISM of galaxies through their
strong winds and high UV fluxes, and are a major source of the heavy
elements enriching the ISM \citep[e.g,][and references
therein]{massey03,zinnecker07}. A large fraction of massive stars are
found in binaries \citep[e.g.,][]{kiminki07}. Eclipsing binaries are of
particular use because they allow us to determine the masses and radii
of the components and the distance to the system. Many young, massive
eclipsing binaries have been found and studied in our Galaxy, the LMC,
and the SMC, primarily in OB associations and young star clusters
\citep[e.g.,][]{bonanos04,peeples07,gonzalez05,hilditch05}. The study of
massive eclipsing binaries beyond the Magellanic clouds has been limited
until very recently, when variability searches using medium-sized
telescopes with wide-field CCD cameras, coupled with spectroscopy using
8-meter class telescopes, have yielded the first systems with accurately
measured masses in M31 \citep{ribas05} and M33 \citep{bonanos06}.

We conducted a deep variability survey of M81 and its dwarf irregular
companion, Holmberg~IX, using the Large Binocular Camera (LBC) 
mounted on the Large Binocular Telescope (LBT),
between January and October 2007. Holmberg~IX is a young dwarf galaxy (age
$\la 200$~Myr), with a stellar population dominated by blue and red
supergiants with no signs of old stars in the red giant branch
\citep{makarova02}. The dwarf may have formed during a recent tidal
interaction between M81 and NGC~2976 \citep[e.g.,][]{boyce01}.  The
gas-phase metal abundance of Holmberg~IX of between 1/8 and 1/3 solar
\citep[e.g.,][]{miller95,makarova02} is consistent with this hypothesis
\citep[e.g.,][]{weilbacher03}. A normal, isolated dwarf on the
luminosity-metallicity relationship would have a metallicity of $\sim
1/20$ solar \citep{lee06}.

In this Letter, we report on the discovery of a $271$~day period,
evolved, massive eclipsing binary in Holmberg~IX using data from the
LBT. The overcontact system is the brightest periodic variable
discovered in our LBT variability survey. It has an out-of-eclipse
magnitude of $V_{max}=20.7$~mag and is located at $\alpha = 09^{\rm
h}57^{\rm m}37\fs 14$, $\delta = +69^{\circ} 02' 11''$ (J2000.0). In
\S~2 we discuss the observations and data reduction, in \S~3 we present
the light curve and the best-fit eclipsing binary model, and note a
similar eclipsing binary in the SMC. In \S~4 we discuss the results and
their possible implications for Type~II supernovae. Throughout this
Letter, we assume the HST Key Project distance to M81 of $\mu=27.80$~mag
($3.6$~Mpc; \citealt{freedman01}) as the distance to Holmberg~IX, and
correct only for a foreground Galactic extinction of $E(B-V)=0.08$~mag
\citep{sfd}.

\section{Observations}

Holmberg~IX was observed as part of a variability survey of the entire
M81 galaxy conducted between January and October~2007 with the LBT 8.4-meter
telescope \citep{hill06}, using the LBC-Blue CCD camera \citep{rag06,gial07} 
during Science Demonstration Time. The survey cadence and depth 
(1~min single exposures, with $\geq 3$ consecutive exposures) are 
optimized to detect and follow-up Cepheid variables with periods between 
10--100 days ($V\la 24$~mag), getting better than 10\% photometry in the 
$B$ and $V$ filters. We obtained 168 $V$-band images
on 24 different nights, and 87 $B$-band images on 13 nights. We coadded
the $B$-band images from each night (usually 3--4) to improve the
signal-to-noise in the combined images. The seeing (FWHM) varied between
$0\farcs 7-2\farcs0$ in $V$ (median $1\farcs 4$), and between $1\farcs 0
- 3\farcs 3$ in $B$-band (median $1\farcs 9$). Our program did not
request especially good image quality for these queue scheduled SDT
observations.

We also observed Holmberg~IX as part of a variability survey of M81
conducted with the 8K~Mosaic imager mounted on the MDM 2.4-meter
telescope. The observations were obtained in 5 one-week runs between
February~2006 and February~2007. All the images were obtained in
$V$-band using 15~min exposures. Due to weather loses and bad seeing, we
ended up using only 36 images from 12 different nights. The typical
seeing was $\sim 1\farcs 1$.

\section{Light Curve}

We used the ISIS difference image analysis package
\citep{alard,hartman04} to obtain the $V$-band light curves of all the
point sources detected in the LBT reference image. The detection of all
point sources and the transformation of difference-flux light curves to
instrumental magnitudes were done using the DAOPHOT/ALLSTAR package
\citep{stetson87,stetson92}. After visual inspection of all the light
curves of variable point-sources selected by standard criteria (rms and
AoV significance; \citealt{hartman07}), we detected \mbox{$\sim 20$}
periodic variables in the field of Holmberg~IX. These include Cepheids
with periods of $10-60$~days and one long-period variable. The analysis
of the Cepheid PL-relation and the distance to Holmberg~IX will be
presented in a future paper (Prieto et~al. 2008, in preparation). The
brightest periodic variable is the peculiar, long-period
($P=270.7$~days) eclipsing binary we discuss here (hereafter V1).

After discovering the binary in the LBT data, we also ran ISIS and
DAOPHOT/ALLSTAR on the MDM data to extract the light curve of the
long-period binary. The variability data from LBT and MDM were
complemented with single-epoch archival imaging of the field obtained
from the SDSS Data Release 6 \citep{sdss} in the $gri$ bands (UT
Nov. 30, 2003), and the HST/ACS Wide Field Camera (GO proposal 10605, PI
E.~Skillman) in the $F555W$ and $F814W$ filters (UT Mar. 23, 2006).  The
high resolution HST/ACS images (${\rm FWHM} \sim 0\farcs 1$,
corresponding to $\sim 2$~pc at the distance of Holmberg~IX) show that
the binary is spatially coincident with a stellar association in the
dwarf galaxy.

\begin{figure*}[!t]
\epsscale{1.0} 
\plotone{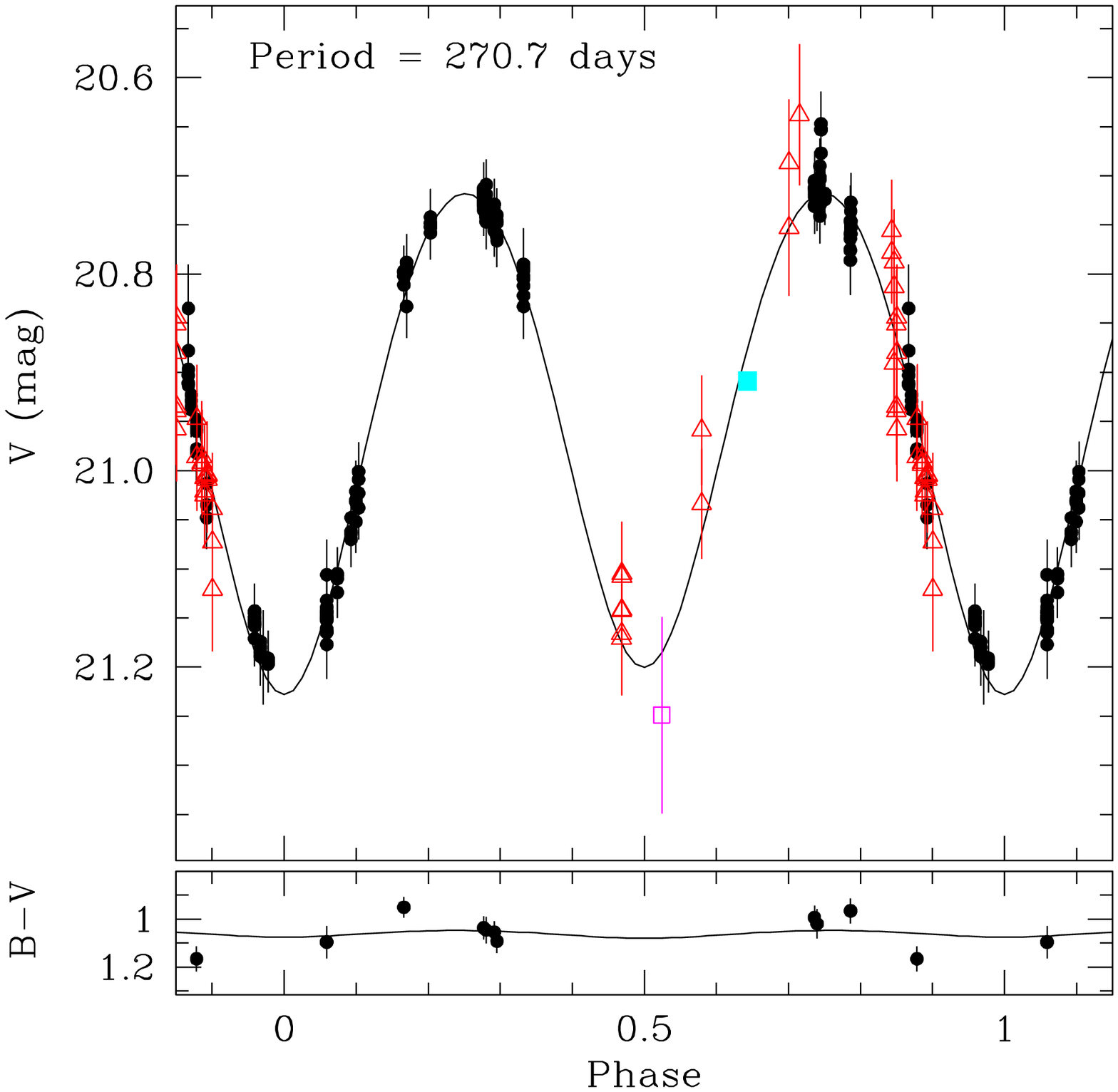}
\caption{Phased $V$-band light curve of the 270.7~day period eclipsing
binary in Holmberg~IX ({\it top panel}) and its $B-V$ color evolution
({\it lower panel}). The different symbols correspond to photometry from
different telescopes: LBT ({\it filled circles}), MDM ({\it open
triangles}), HST/ACS ({\it filled square}) and SDSS ({\it open
square}). The {\it solid line} shows an overcontact eclipsing binary
model that best fits the light $V$ light curve. \label{fig1}}
\end{figure*}

Figure~\ref{fig1} shows the phased $V$-band light curve and $B-V$ color
curve of the eclipsing binary system. We include all the LBT, MDM, SDSS
and HST/ACS $V$-band photometry. The LBT and MDM photometry have been
calibrated using SDSS photometry of several relatively bright ($r \la
21.0$~mag) and unsaturated stars in the field, transforming the $gr$
magnitudes to standard $BV$ magnitudes with the transformations
presented in \citet{ivezic07}. The rms deviations of the absolute
calibration are $0.02-0.03$~mag for LBT-$BV$, and $\sim 0.05$~mag for
MDM-$V$. The SDSS $gri$ photometry of the binary was extracted in the
same way as for LBT and MDM, using the DAOPHOT/ALLSTAR package to obtain
instrumental magnitudes calibrated using absolute photometry of the
bright stars in the field. Our $g$ and $r$ magnitudes of the binary from
the SDSS data are 0.2 and 0.5~mag brighter, respectively, than the
magnitudes reported in the SDSS-DR6 catalog, while the $i$-band
magnitude agrees at the 1\% level. We think this is due to problems in
the SDSS photometry for faint sources in a crowded field
\citep[e.g.,][]{smolcic07}. The details of the HST photometry can be
found in Weisz et al. (2008, in preparation). 

\begin{figure*}[!t]
\epsscale{1.0}
\plotone{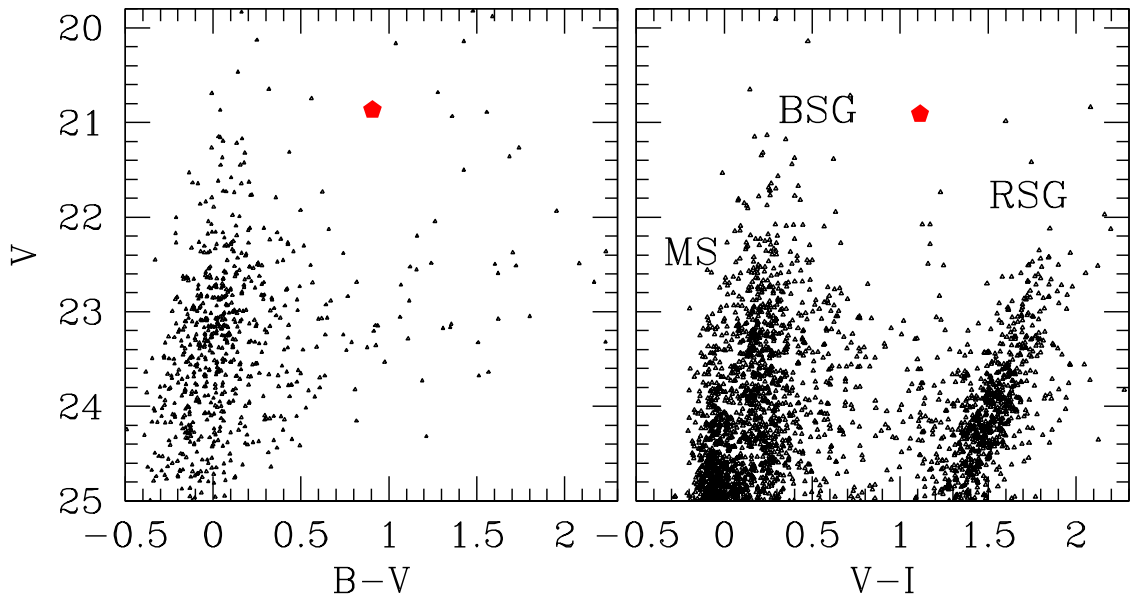}
\caption{Color Magnitude Diagrams of stars in the field of Holmberg~IX
obtained from the LBT-$BV$ reference images ({\it left panel}) and
HST/ACS-$VI$ single-epoch observations ({\it right panel}). The ACS CMD
shows well-defined stellar sequences for the main sequence (MS) and the
evolved blue (BSG) and red supergiants (RSG).  The eclipsing binary
({\it pentagon}) lies between the blue and red supergiant
sequences. \label{fig2}}
\end{figure*}

In Figure~\ref{fig2} we show the position of the binary in the color
magnitude diagrams (CMD), obtained from calibrated LBT and HST/ACS
photometry. The CMDs show the well-populated blue and red supergiant
sequences in Holmberg~IX. The binary is among the most luminous stars in
this dwarf galaxy, with $M_{V} \sim -7.1$~mag, and it has clearly
evolved from the main-sequence. With such a high intrinsic luminosity,
the binary is bound to be massive. After correcting for Galactic
foreground extinction, the $B-V$ and $V-I$ colors are consistent with an
effective temperature of $T_{\rm eff} = 4800\pm 150$~K
\citep{houdashelt00}. Both components seem to be G-type yellow
supergiants given the equal depths of the eclipses and the lack of color
variations (see Figure~\ref{fig1}).

We used the eclipsing binary model-fitting program
NIGHTFALL\footnote{\scriptsize{{\tt
http://www.hs.uni-hamburg.de/DE/Ins/Per/Wichmann/Nightfall.html}}} to
model the $V$-band light curve. As shown in Figure~\ref{fig1}, we obtain
a good fit to the light curve with an overcontact configuration where
both stars are overflowing their Roche lobes. We fixed the effective
temperature of the primary at $T_{1}=4800$~K obtained from the
colors. We assumed equal masses for the stars, a linear limb-darkening
law, circular orbits, and synchronous rotation.  We fit for four
parameters: the Roche lobe filling factors, the inclination, and the
temperature of the secondary. The time of the primary eclipse and the
period were determined externally and were fixed for these fits. The
main parameters of the binary are listed in Table~\ref{tab1}. The light
curve shows a hint of the \citet{oconnell} effect, in which the maxima
(out-of-eclipse regions) show a difference in brightness
\citep[e.g.,][]{pilecki07}.

\begin{figure*}[!t]
\epsscale{1.0}
\plotone{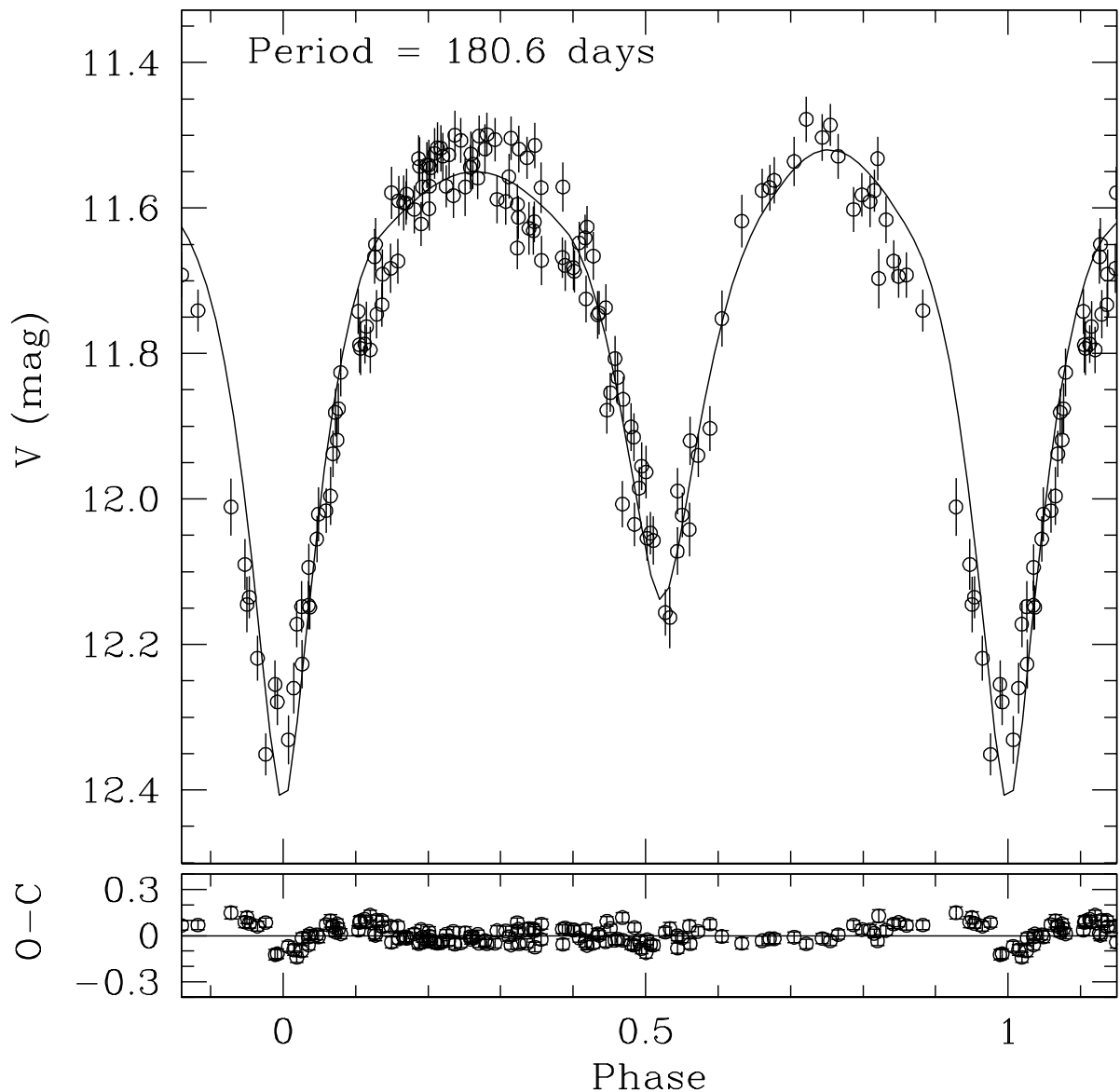}
\caption{Phased $V$-band light curve of the long-period, evolved
eclipsing binary SMC~R47 obtained from the ASAS catalog
\citep{asas}. The star was classified spectroscopically as an F0
supergiant by \citet{humphreys83}. The {\it solid line} shows the
contact eclipsing binary model that best fits the light curve. The {\it
lower panel} shows the residuals of the fit
(observed-calculated). \label{fig3}}
\end{figure*}

We searched the literature for other examples of evolved, massive
eclipsing binaries in the yellow supergiant phase and found
none\footnote{Note, however, that a possible Galactic counterpart is the
contact binary BM Cas \citep{fernie97}, composed by an A7~Iab supergiant
($M_{V}\simeq -6.3$) and a late-type giant.}. We also searched the
available catalogs of eclipsing binaries in the LMC and SMC. The MACHO
catalog of eclipsing binaries in the LMC \citep{derekas07} contains 25
contact systems with red colors (i.e., evolved), $(V-R) > 0.5$~mag, and
periods $>200$~days. However, these systems have absolute magnitudes
$M_{V} \ga -4$ ($V\ga 14.5$~mag), that are $\sim 3$~mag fainter than the
yellow supergiant eclipsing binary in
Holmberg~IX\footnote{\citet{mennickent06} obtained spectroscopy of 17
``peculiar'' periodic variables in the SMC from the OGLE database, and
found a 184~day period eclipsing binary composed by two yellow
supergiants (F5Ie + G5-K0I). However, this system is $\sim 2$~mag
fainter than the binary in Holmberg~IX.}. The All Sky Automated Survey
(ASAS; \citealt{asas}) contains complete Southern sky coverage for
$V<15$~mag. To our surprise, we found in the ASAS catalog a luminous
($M_{V} \sim -7.5$, $V_{max} \sim 11.5$~mag), 181~day period contact
eclipsing binary in the SMC. The star, SMC~R47 ($\alpha = 01^{\rm
h}29^{\rm m}17\fs 26$, $\delta = -72^{\circ} 43' 20.2''$), had been
spectroscopically classified as an FO supergiant ($T_{\rm eff} \simeq
7500$~K) with emission lines by \citet{humphreys83}.  The ASAS $V$-band
light curve of SMC~R47, obtained between December 2002 and June 2006,
and the fit obtained with NIGHTFALL are shown in Figure~\ref{fig3}. The
best-fit eclipsing binary model requires a contact configuration, with
non-zero eccentricity to account for the difference in timing between
the eclipses. Even though we selected a clean part of the full ASAS
light curve, there seems to be intrinsic variability from the binary
components. The main parameters of this eclipsing binary are in
Table~\ref{tab1}. While substantially hotter than the Holmberg~IX
binary, it does not lie on the SMC blue supergiant sequence
\citep{grieve86}.

\section{Discussion and Conclusions}

An eclipsing binary is the best explanation for the light curve of the
brightest variable we have discovered in our LBT variability survey of
the dwarf irregular companion of M81 Holmberg~IX. The other possible
explanation for the periodic variability of V1 is a long-period
($P=135$~days) Cepheid. Such long-period Cepheids ($P>100$~days) have
been observed in dwarf galaxies like the LMC and SMC
\citep[e.g.,][]{freedman85}, NGC~55 \citep[ e.g.,][]{pietrzynski06},
NGC~300 \citep[e.g.,][]{pietrzynski02}, NGC~6822
\citep[e.g.,][]{pietrzynski04}, IC~1613 \citep[e.g.,][]{antonello99},
and I~Zw~18 \citep{aloisi07}. The magnitude of V1 is consitent with the
magnitude of a Cepheid with $P=135$~days ($M_{V}\simeq -7.0$),
extrapolating the period-luminosity relationship of
\citet{fouque07}. However, while a few of these long period Cepheids
have quasi-sinusoidal light curves that are nearly symmetric under a
time reversal, they all have larger amplitudes in bluer bands ($B$
amplitudes $1.3-1.6$ times $V$ amplitudes) due to the changes in the
effective temperature as the star pulsates
\citep[e.g.,][]{freedman85,madore91}. Spectroscopy of V1, while
challenging, would eliminate any remaining ambiguities in classifying
this system.
 
We can safely rule out the possibility that the eclipsing binary is in
our Galaxy. Using the period, estimated effective temperature, magnitude
at maximum, and assuming that the stars are of similar size in a contact
configuration, we can use Kepler's law and their total surface
brightness to estimate a distance to the system of $D \simeq
(M_{total}/\Msun)^{1/3}$~Mpc, where $M_{total} = M_{1} + M_{2}$
\citep[e.g.,][]{gaposchkin62}. Conversely, we can estimate the total
mass of the binary system by assuming the distance, \mbox{$M_{total}
\simeq 45\,(D/3.6\,{\rm Mpc})^{3}\,\Msun$}. This should be taken as a
rough estimate because of the overly simple model $-$ to accurately
constrain the total mass of the system, and its components, we need
radial velocity measurements. Another piece of evidence that puts the
binary system in Holmberg~IX is its spatial coincidence with a stellar
overdensity in the dwarf, observed in the HST/ACS images.

\begin{figure*}[!t]
\epsscale{1.0}
\plotone{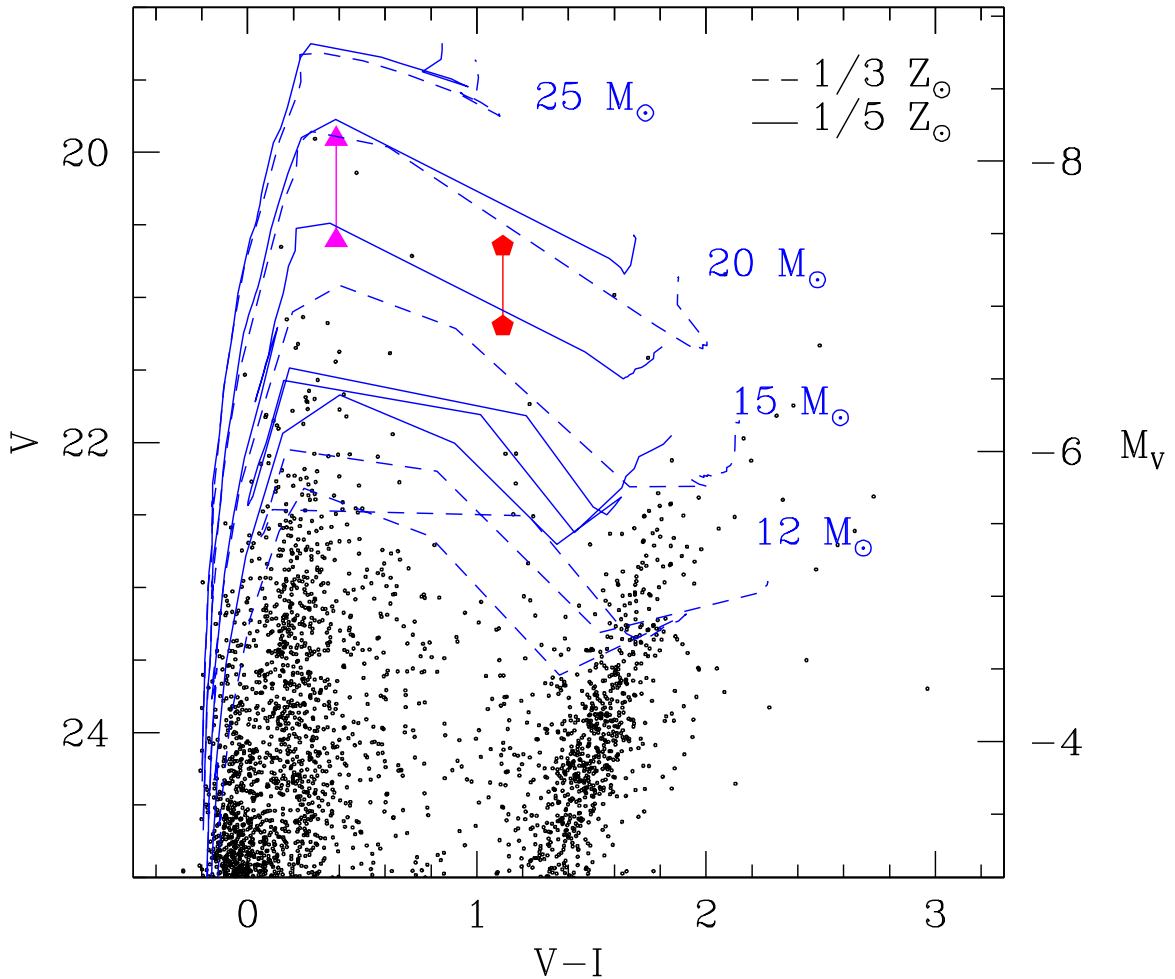}
\caption{CMD of Holmberg~IX from HST/ACS $V$ and $I$ photometry. The
connected {\it filled symbols} show the position of the evolved
eclipsing binaries in Holmberg~IX ({\it filled pentagons}) and the SMC
({\it filled triangles}) at maximum and minimum. The lines show
evolutionary tracks with extended mass-loss from the Geneva group
\citep{geneva} for single stars with masses between 12--25~$\Msun$,
assuming two different metallicities: 1/3 solar ({\it dashed}) and 1/5
solar ({\it solid}). We use a distance modulus of $\mu = 27.80$~mag to
Holmberg~IX, and Galactic color-excess $E(B-V) = 0.08$~mag, to put the
evolutionary tracks in the diagram. \label{fig4}}
\end{figure*}

We expected that such systems were rare\footnote{While the relative
numbers of eclipsing binaries is a much more complicated problem, we
note that the relative abundances of red, blue and yellow supergiants is
4:13:1 for the Geneva evolutionary track \citep{geneva} of a single,
non-rotating star with $M=15\,\Msun$ and $Z=0.004$.}, but were surprised
to find none in the literature. However, we found a similar eclipsing
binary system in the SMC (SMC~R47) searching through the ASAS
catalog. From the absolute magnitudes of both binaries and their colors,
we estimate that at least one of the stars in each binary is $\sim
15-20\,\Msun$ (main-sequence age $\sim 10-15$~Myr) using the
evolutionary tracks for single stars of \citet{geneva} (see
Figure~\ref{fig4}). 

The stellar evolutionary path of stars of a given mass in binary systems
can differ significantly from their evolution in isolation
\citep[e.g.,][]{paczynski71}. In particular, binary interactions through
mass loss, mass accretion, or common-envelope evolution, play a very
important role in the pre-supernova evolution
\citep[e.g.,][]{podsiadlowski92}. Most of the massive stars with masses
$30\,\Msun \ga M \ga 8\,\Msun$ are expected to explode as supernova when
they are in the red supergiant stage, with a small contribution from
blue supergiants (e.g., SN~1987A; \citealt{west87}). Surprisingly,
\citet{li05} identified the progenitor of the Type~IIP supernova 2004et
in pre-explosion archival images and determined that it was a yellow
supergiant with a main-sequence mass of $\sim 15\,\Msun$. Also, the
position in the CMD of the likely progenitor of the Type~IIP supernova
2006ov (see Fig.~10 in \citealt{li07}) is remarkably similar to the
position of the eclipsing binary in Holmberg~IX (see Figure~\ref{fig4}).

We propose that the binary we discovered in Holmberg~IX and the binary
found in the SMC\footnote{A possible earlier stage of these yellow
supergiant binaries might be represented by the B-supergiant pair HD
1383 \citep{boyajian06}.} are the kind of progenitor objects of
supernovae like SN~2004et and SN~2006ov that appeared to be the
explosions of yellow supergiants. A close binary provides a natural
means of slowing the transition from blue to red, allowing the star to
evolve and then explode as a yellow supergiant. As the more massive star
evolves and expands, the Roche lobe limits the size of the star forcing
it to have a surface temperature set by the uncoupled core luminosity
and the size of the Roche lobe. It can expand further and have a cooler
envelope only by becoming a common envelope system, which should only
occur as the secondary evolves to fill its Roche lobe. This delayed
temperature evolution allows the core to reach SN~II conditions without
a red envelope.

\acknowledgments 

We are grateful to the referee, Douglas~Gies, for his helpful comments
and suggestions. The authors thank the LBT Science Demonstration Time
(SDT) team for assembling and executing the SDT program. We also thank
the LBC team and the LBTO staff for their kind assistance. We are
grateful to Alceste Bonanos, Marc Pinsonneault, Lev Yungelson and Stan
Woosley for helpful discussions and comments on earlier versions of this
draft. We thank G.~Pojmanski for making the excellent ASAS data publicly
available. JLP and KZS acknowledge support from NSF through grant
AST-0707982.

\newpage 

\begin{deluxetable}{lll}
\tablewidth{-1pt}
\tablecaption{Best-fit Binary Model Parameters. \label{tab1} }
\tablehead{
\colhead{Parameter} &
\colhead{Holmberg~IX~V1} &
\colhead{SMC~R47}
}
\startdata
Period, $P$                                    & $270.7 \pm 2.3$~days       &  $181.58 \pm 0.16$~days   \\
Time of primary eclipse, $T_{prim}$            & $2454186.0 \pm 0.6$        &  $2452073.1 \pm 0.2$     \\
Inclination, $i$                               & $55.7\degr \pm 0.6 \degr$  &  $82.2\degr \pm 0.2 \degr$   \\
Primary temperature, $T_{1}$                   & $4800 \pm 150$~K           &  $7500 \pm 100$~K        \\
Temperature ratio, $T_{2}/T_{1}$               & $1.05 \pm 0.05$            &  $1.17 \pm 0.02$   \\
Eccentricity, $e$                              & 0.00                       &  $0.039 \pm 0.002$\tablenotemark{a} \\
Roche Lobe Filling factors\tablenotemark{b}    & $1.23 \pm 0.02$            &  $1.02 \pm 0.02$ \\
Semi-major axis, $a$\tablenotemark{c}          & 547~$R_{\odot}$ &  418~$R_{\odot}$    \\ 
\enddata
\tablecomments{The mass ratio was fixed at $q=1$ for fitting both light curves.}
\tablenotetext{a}{A non-zero eccentricity is required to fit the difference in timing between the primary and secondary eclipses. The best-fit longitude of the periastron is $w=168.6\degr \pm 1.1\degr$.}
\tablenotetext{b}{Ratio of stellar to Roche lobe polar radius for each star.}
\tablenotetext{c}{Separation between the stars assuming a total mass of $30\,\Msun$ for each system.}
\end{deluxetable}


\begin{thebibliography}{}
\bibitem[Alard(2000)]{alard} Alard, C. 2000, \aap, 144, 363
\bibitem[Aloisi et al.(2007)]{aloisi07} Aloisi, A., et~al.\ 2007, \apjl, 667, L151
\bibitem[Adelman-McCarthy et al.(2007)]{sdss} Adelman-McCarthy, J., et al. 2007, \apjs, submitted (arxiv:0707.3413) 
\bibitem[Antonello et al.(1999)]{antonello99} Antonello, E., et~al.\ 1999, \aap, 349, 55
\bibitem[Bonanos et al.(2004)]{bonanos04} Bonanos, A.~Z., et~al.\ 2004, \apjl, 611, L33
\bibitem[Bonanos et al.(2006)]{bonanos06} Bonanos, A.~Z., et~al.\ 2006, \apj, 652, 313
\bibitem[Boyajian et al.(2006)]{boyajian06} Boyajian, T.~S., et~al.\ 2006, \apj, 646, 1209
\bibitem[Boyce et al.(2001)]{boyce01} Boyce, P.~J., et~al.\ 2001, \apjl, 560, L127
\bibitem[Derekas et al.(2007)]{derekas07} Derekas, A., Kiss, L.~L., \& Bedding T.~R.\ 2007, \apj,  663, 249
\bibitem[Fernie \& Evans(1997)]{fernie97} Fernie, J.~D., \& Evans, N.~R.\ 1997, \pasp, 109, 541 
\bibitem[Fouque et al.(2007)]{fouque07} Fouque, P., et~al.\ 2007, \aap, in press (arxiv:0709.3255)
\bibitem[Freedman et al.(1985)]{freedman85} Freedman, W.~L., Grieve, G.~R., \& Madore, B.~F.\ 1985, \apjs, 59, 311
\bibitem[Freedman et al.(2001)]{freedman01} Freedman, W.~L., et~al.\ 2001, \apj, 553, 47
\bibitem[Gaposchkin(1962)]{gaposchkin62} Gaposchkin, S.\ 1962, \aj, 67, 358
\bibitem[Giallongo et al.(2007)]{gial07} Giallongo, E., Ragazzoni, R., Grazian, A. et al. 2007 in preparation
\bibitem[Gonz{\'a}lez et al.(2005)]{gonzalez05} Gonz{\'a}lez, J.~F., et~al.\ 2005, \apj, 624, 946
\bibitem[Grieve \& Madore(1986)]{grieve86} Grieve, G.~R., \& Madore, B.~F.\ 1986, \apjs, 62, 427
\bibitem[Hartman et al.(2004)]{hartman04} Hartman, J.~D., et~al.\ 2004, \aj, 128, 1761 
\bibitem[Hartman et al.(2007)]{hartman07} Hartman, J.~D., et~al.\ 2007, \apj, submitted (arxiv:0709.3484) 
\bibitem[Hilditch et al.(2005)]{hilditch05} Hilditch, R.~W., Howarth, I.~D., \& Harries, T.~J.\ 2005, \mnras, 357, 304
\bibitem[Hill et al.(2006)]{hill06} Hill, J.~M., Green, R.~F., \& Slagle, J.~H.\ 2006, \procspie, 6267, 62670Y
\bibitem[Houdashelt et al.(2000)]{houdashelt00} Houdashelt, M.~L., Bell, R.~A., \& Sweigart, A.~V.\ 2000, \aj, 119, 1448
\bibitem[Humphreys(1983)]{humphreys83} Humphreys, R.~M.\ 1983, \apj, 265, 176
\bibitem[Ivezi{\'c} et~al.(2007)]{ivezic07}  Ivezi{\'c}, {\v Z}., et~al.\ 2007, \aj, 134, 973
\bibitem[Kiminki et~al.(2007)]{kiminki07} Kiminki, D.~C., et~al.\ 2007, \apj, 664, 1102
\bibitem[Lee et~al.(2006)]{lee06} Lee, H., et~al. 2006, \apj, 647, 970
\bibitem[Lejeune \& Schaerer(2001)]{geneva} Lejeune, T., \& Schaerer, D.\ 2001, \aap, 366, 538
\bibitem[Li et al.(2005)]{li05} Li, W., et~al.\ 2005, \pasp, 117, 121
\bibitem[Li et al.(2007)]{li07} Li, W., et~al.\ 2007, \apj, 661, 1013 
\bibitem[Madore \& Freedman(1991)]{madore91} Madore, B.~F., \& Freedman, W.~L.\ 1991, \pasp, 103, 933
\bibitem[Makarova et al.(2002)]{makarova02} Makarova, L.~N., et~al.\ 2002, \aap, 396, 473
\bibitem[Massey(2003)]{massey03} Massey, P.\ 2003, \araa, 41, 15
\bibitem[Mennickent et al.(2006)]{mennickent06} Mennickent, R.~E., et~al.\ 2006, RMxAC, 26, 74
\bibitem[Miller(1995)]{miller95} Miller, B.~W.\ 1995, \apjl, 446, L75 
\bibitem[O'Connell(1951)]{oconnell} O'Connell, D.~J.~K.\ 1951, Publications of the Riverview College Observatory, 2, 85
\bibitem[Paczy{\'n}ski(1971)]{paczynski71} Paczy{\'n}ski, B.\ 1971, \araa, 9, 183
\bibitem[Peeples et al.(2007)]{peeples07} Peeples, M.~S., et~al.\ 2007, \apjl, 654, L61 
\bibitem[Pietrzy{\'n}ski et al.(2006)]{pietrzynski06} Pietrzy{\'n}ski, G., et~al.\ 2006, \apj, 132, 2556
\bibitem[Pietrzy{\'n}ski et al.(2004)]{pietrzynski04} Pietrzy{\'n}ski, G., et~al.\ 2004, \apj, 128, 2815
\bibitem[Pietrzy{\'n}ski et al.(2002)]{pietrzynski02} Pietrzy{\'n}ski, G., et~al.\ 2002, \apj, 123, 789
\bibitem[Pilecki et~al.(2007)]{pilecki07} Pilecki, B., Fabrycky, D., \& Poleski, R.\ 2007, \mnras, 378, 757
\bibitem[Podsiadlowski et al.(1992)]{podsiadlowski92} Podsiadlowski, P., Joss, P.~C., \& Hsu, J.~J.~L\ 1992, \apj, 391, 246
\bibitem[Pojmanski(2002)]{asas} Pojmanski, G. 2002, \actaa, 52, 397
\bibitem[Ragazzoni et al.(2006)]{rag06} Ragazzoni, R., et al.\ 2006, \procspie, 6267, 626710
\bibitem[Ribas et al.(2005)]{ribas05} Ribas, I., et al.\ 2005, \apjl, 635, L37
\bibitem[Schlegel et al.(1998)]{sfd} Schlegel, D.~J., Finkbeiner, D.~P., \& Davis, M.\ 1998, \apj, 500, 525
\bibitem[Smol{\v c}i{\'c} et al.(2007)]{smolcic07} Smol{\v c}i{\'c}, V., et al.\ 2007, \aj, 134, 1901
\bibitem[Stetson(1987)]{stetson87} Stetson, P.~B.\ 1987, \pasp, 99, 191
\bibitem[Stetson(1992)]{stetson92} Stetson, P.~B.\ 1992, \jrasc. 86, 71
\bibitem[Weilbacher et al.(2003)]{weilbacher03} Weilbacher, P.~M., et~al.\ 2003, \aap, 397, 545
\bibitem[West et al.(1987)]{west87} West, R.~M., et~al.\ 1987, \aap, 177, L1
\bibitem[Zinnecker \& Yorke(2007)]{zinnecker07} Zinnecker, H., \& Yorke, H.~W.\ 2007, \araa, 45, 481

\end{thebibliography}
\end{document}